\documentclass[aps,preprint]{revtex4}%
\usepackage{amsfonts}
\usepackage{amsmath}
\usepackage{amssymb}
\usepackage{graphicx}%
\providecommand{\U}[1]{\protect\rule{.1in}{.1in}}

\begin{document}
\title{Pressure-induced superconductivity in CaLi$_{2}$}
\author{T. Matsuoka,$^{\ast}$ M. Debessai, J. J. Hamlin, A. K. Gangopadhyay, and J. S. Schilling}
\affiliation{Department of Physics, Washington University, CB 1105, One Brookings Drive,
St. Louis, MO 63130, USA}
\author{K. Shimizu}
\affiliation{KYOKUGEN, Center for Quantum Science and Technology under Extreme Conditions,
Osaka University, 1-3 Machikaneyama, Toyonaka, Osaka 560-8531, Japan}

\begin{abstract}
A search for superconductivity has been carried out on the hexagonal polymorph
of Laves-phase CaLi$_{2}$, a compound for which Feng, Ashcroft, and Hoffmann
predict highly anomalous behavior under pressure. No superconductivity is
observed above 1.10 K at ambient pressure. However, high-pressure ac
susceptibility and electrical resistivity studies to 81 GPa reveal bulk
superconductivity in CaLi$_{2}$ at temperatures as high as 13 K. The
normal-state resistivity displays a dramatic increase with pressure.

\vspace{7.3cm}\noindent$^{\ast}$Permanent address: \ KYOKUGEN, Center for
Quantum Science and Technology under Extreme Conditions, Osaka University, 1-3
Machikaneyama, Toyonaka, Osaka 560-8531, Japan

\end{abstract}
\date{February 20, 2008}
\maketitle

Li and Ca are both `simple' metals with one (1$s$) and two (2$s$) conduction
electrons per atom, respectively. Recently, Tuoriniemi \textit{et al.}
\cite{tuoriniemi} reported superconductivity in Li at extremely low
temperatures ($T_{c}\approx$ 0.4 mK). If Li is subjected to increasing
pressure, a superconducting transition abruptly appears at $T_{c}\approx$ 5 K
for 20 GPa and climbs steeply, passing through a maximum at 14 - 20 K for 30
GPa \cite{shimizu10,struzhkin10,deemyad10}. In addition, the electrical
resistivity of Li increases sharply with pressure \cite{shimizu10,lin1}. These
results are in agreement with the prediction of Neaton and Ashcroft
\cite{neaton1} that Li's electronic properties should become highly anomalous
at extreme pressures. On the other hand, Ca is not known to superconduct at
ambient pressure, but becomes superconducting at $\sim$ 1.2 K under 50 GPa
\cite{okada1}, reaching 25 K at 160 GPa (1.6 Mbar) \cite{yabuuchi10}, a record
value of $T_{c}$ for an elemental superconductor. The electrical resistivity
of Ca also increases strongly with pressure \cite{yabuuchi10}. It is notable
that at these extreme pressures the conduction bandwidths in Li \cite{neaton1}%
, Na \cite{neaton2}, and Ca \cite{lei1} are calculated to \textit{decrease}
under pressure, a counterintuitive result.

Recently, Feng \textit{et al.} \cite{feng1} pose the interesting question
whether the anomalous properties predicted and found for Li and Ca under
pressure carry over to the only known binary compound containing these two
elements, the Laves phase CaLi$_{2}$. Detailed electronic structure
calculations by these authors for both hexagonal and cubic polymorphs of
CaLi$_{2}$ lead them to predict that \textquotedblleft the elevated density of
states at the Fermi level, coupled with the expected high dynamical scale of
Li as well as the possibility of favorable interlayer phonons, points to
potential superconductivity of CaLi$_{2}$ under pressure\textquotedblright. As
for Li and Ca, the conduction bandwidth is predicted to decrease under high
compression, narrowing at 139 GPa to only one-third the free-electron value
\cite{feng1}. The anomalous electronic properties in Li, Ca, and CaLi$_{2}$
under extreme pressures\ arise from the fact that as the ionic cores of the
constituent elements approach each other and begin to overlap, the conduction
electrons, which are excluded from these ion cores, forfeit their
nearly-free-electron character\ as they are forced into cramped, low-symmetry
interstitial sites \cite{neaton1,neaton2,feng1}. These arguments are quite
general and should apply to a wide range of nominally free-electron elements,
alloys, and compounds \cite{neaton2}. Since CaLi$_{2} $ serves as a test case
for these ideas, high-pressure measurements on this material are of particular
importance. To our knowledge, CaLi$_{2}$ has not yet been tested for
superconductivity at any pressure, ambient or otherwise.

In this paper we report measurements of the temperature-dependent electrical
resistivity and ac susceptibility of hexagonal CaLi$_{2}$ at both ambient and
high pressure to 81 GPa in a diamond-anvil cell (DAC). At ambient pressure no
superconductivity is found above 1.10 K. For pressures above 11 GPa a
superconducting transition is observed in both the electrical resistivity and
ac susceptibility where $T_{c}$ increases with pressure and passes through a
maximum at 11-13 K near 40 GPa. In the normal state the electrical resistivity
shows a remarkably large increase with pressure over the measured temperature
range 2 - 300 K.

The starting materials for the synthesis of CaLi$_{2}$ are pieces of 99.98\%
Ca from Alfa Aesar and 99.99\% Li rod from ESPI Metals. The first CaLi$_{2}$
sample (sample A) was prepared by melting together stoichiometric amounts of
Ca and Li in a stainless steel crucible placed on a hot plate in an Ar-gas
glovebox. A 13 $\mu$m thick Ta foil was placed under the sample in the
crucible and a thin Ta strip was used to stir the molten sample for
approximately 30 minutes at a temperature somewhat above the melting
temperature (235$^{\circ}$C) of CaLi$_{2}$. The residual resistivity ratio was
found to be $\rho($300 K$)/\rho($2 K$)\simeq170,$ a relatively high value for
an intermetallic compound, which speaks for its phase purity. A second
CaLi$_{2}$ sample (sample B) was prepared as sample A, but was then wrapped in
a Ta foil, sealed in a quartz tube filled with Ar, and annealed for 24 hours
in a box furnace at a temperature slightly below the melting temperature.
Scanning electron microscopy (Hitachi S-4500) revealed the presence of a small
amount of impurity phase (%
$<$
5 vol\%) in sample A. The annealing procedure significantly reduced the amount
of impurity phase in sample B to below 1 vol\%.

In preparation for X-ray powder diffraction studies, an agate mortar and
pestle in the Ar glovebox was used to grind the brittle CaLi$_{2}$ sample to a
powder which was filled together with a Si marker into glass capillary tubes
(0.9 mm O.D. $\times$ 0.01 mm wall) and sealed shut under Ar gas. Five such
capillary tubes were placed in a Rigaku Geigerflex D/max-B X-ray
diffractometer utilizing Cu-K$_{\alpha}$ radiation. The powder diffraction
pattern of both samples indicates single-phase material and confirms the
hexagonal unit cell with $a=6.293(1)$ \AA \ and \ $c=10.236(1)$\ \AA \ for
sample A and $a=6.287(1)$ \AA \ and \ $c=10.233(1)$\ \AA \ for sample B, in
reasonable agreement with published values $a=6.2899(5)$ \AA \ and
\ $c=10.268(1)$\ \AA \ \cite{fischer1}. To search for possible impurity phases
of unreacted Li or Ca, high resolution X-ray diffraction studies were carried
out on sample B for those strong diffraction peaks with indices for Li (110)
and for Ca (200), (220), and (311) which are well separated from the peaks of
CaLi$_{2}$. Using the above diffractometer, no trace of any of these
diffraction peaks could be detected within experimental resolution; this
allows the estimate that the crystalline impurity concentration is below 1.7\%
for Li and 0.3\% for Ca. From very high resolution synchrotron radiation
studies on this sample, the estimate of the impurity limit could be further
reduced to $\leq0.3\%$ for Li and $\leq0.1\%$ for Ca. Full details of the
X-ray diffraction and scanning electron microscopy experiments will be
published elsewhere \cite{matsuoka3}.

At ambient pressure no superconductivity could be detected in CaLi$_{2}$
either in the electrical resistivity to 2 K (samples A and B) nor in sensitive
ac susceptibility measurements to 1.10 K (sample B) where even a 0.03\%
shielding effect would have been detected. High pressure experiments were
carried out using two different DACs, one (type 1) designed by one of the
authors (JSS) \cite{schilling5} and the other (type 2) brought by a second
author (TM) from Osaka University. The type-1 \ DAC uses two opposing
1/6-carat, type Ia diamond anvils with 0.3 mm or 0.5 mm diameter culets,
whereas the culet diameter in the type-2 \ DAC, with 1/4-carat synthetic type
Ib diamonds, is 0.3 mm. Re and W - 25 at.\% Re gaskets (thickness $\sim$ 0.25
mm) are used in the resistivity and ac susceptibility studies, respectively.
The superconductivity of the latter gaskets at 5 K restricts the search for
superconductivity in CaLi$_{2}$ in the present ac susceptibility studies to
temperatures above 5 K. Tiny ruby spheres \cite{chervin} are placed on the
sample to allow the pressure determination using the revised ruby calibration
of Chijioke \textit{et al.} \cite{chijioke} with resolution $\pm$ 0.2 GPa.

A standard four-point ac electrical resistivity technique is used at 13 Hz
frequency with a Stanford Research SR830 digital lock-in amplifier. The sample
is approximately 100 $\mu$m\ in diameter and 10 $\mu$m\ thick, the voltage
leads being approximately 20-30 $\mu$m\ apart. No pressure medium is used so
that the pressure applied to the sample is best characterized as
\textquotedblleft nonhydrostatic\textquotedblright. As a result of the
pressure gradient across the sample and the sizeable sheear stress on the ruby
spheres, the width of the R$_{1}$ ruby fluoresence line \cite{chai1} is as
large as $\pm$ 15 GPa (see Fig.~3 below). In the type-1 DAC the ac
susceptibility is measured using two compensated primary/secondary coil
systems, one around the diamond anvils and the other just outside. For signal
detection the SR830 lock-in amplifier is used with an applied field of 3 Oe
r.m.s. at 1023 Hz. As in the resistivity measurements, no pressure medium is
used. The width of the ruby R$_{1}$ line is a good deal narrower than in the
resistivity studies, as seen in Fig.~3. Further details of the DAC techniques
used in the electrical resistivity \cite{matsuoka1} and ac susceptibility
\cite{matsuoka3,schilling5,hamlin5} measurements are given elsewhere.

The results of the present electrical resistivity measurements on sample A are
shown in Fig.~1. The abrupt drop in the resistance below 15 K for pressures at
or above 11 GPa is the signature of a superconducting transition. As the
pressure is increased above 11 GPa, the transition is seen to shift to higher
temperatures, reaching a maximum value near 13 K for $P\approx40$ GPa. Except
for the measurement at 52 GPa, the normal-state resistivity is seen to rapidly
increase with pressure over the entire temperature range up to the highest
pressure measured (81 GPa). That this rapid resistivity increase is quite
reversible, and thus not mainly due to the addition of lattice defects as the
sample is plastically deformed by the nonhydrostatic pressure, is seen from
the final set of data at 46 GPa (dashed line in Fig.~1) where the pressure was
reduced from 81 GPa. This conclusion is also supported by purely hydrostatic
pressure studies to 0.76 GPa in our He-gas pressure system where the
resistivity is found to increase rapidly and reversibly over the measured
temperature range 14 - 300 K, the rate of increase at ambient temperature
being +13.3(3) \%/GPa. This large pressure-induced increase in the
normal-state resistivity with pressure is analogous to that found in elemental
Li \cite{shimizu10,lin1} and Ca \cite{yabuuchi10} and corroborates the
predictions of Feng \textit{et al.} \cite{feng1} that the electronic
properties of CaLi$_{2}$ should become highly anomalous under pressure. Full
details of the resistivity studies will be given elsewhere \cite{matsuoka3}.

A much superior test for bulk superconductivity than the electrical
resistivity is to search for the onset of strong diamagnetic shielding in the
magnetic susceptibility. In Fig.~2 we plot the real part of the ac
susceptibility versus temperature for monotonically increasing pressure.
Whereas no diamagnetic transition is visible above 5 K at 25 GPa, at 35 GPa
strong diamagnetic shielding consistent with 70\% - 100\% flux expulsion is
seen below 12 K. At somewhat higher pressures the superconducting transition
temperature $T_{c}$ passes through a maximum before falling off rapidly to
temperatures below 5 K for $P\approx$ 54 GPa. As expected for a
superconducting transition, the diamagnetic signal for 39 GPa is seen in the
inset to Fig.~2 to shift by 0.62 K to lower temperatures if a dc magnetic
field of 500 Oe is applied. The same field shifts $T_{c}$ down by 0.54 K at 35
GPa and by 0.48 K at 43 GPa. The small magnitude of these field-induced shifts
gives evidence that CaLi$_{2}$ is a type II superconductor.

In Fig.~3 the measured values of $T_{c}$ are plotted versus pressure for both
the resistivity and ac susceptibility measurements. Here $T_{c}$ is defined
from the transition midpoint, rather than from the transition onset
\cite{experiment1}. For applied pressures above 11 GPa, $T_{c}(P)$ from both
sets of measurements is seen to initially shift under pressure to higher
temperatures but then to pass through a maximum near 12 K for $P\approx40$
GPa. The differences in the $T_{c}(P)$ dependences in resistivity and
susceptibility measurements likely arise from the appreciable pressure
gradients across the sample and/or shear stress effects in these
nonhydrostatic experiments. The susceptibility measurement should yield the
more intrinsic $T_{c}(P)$ dependence since it is a superior measurement of
bulk superconductivity. The rather sharp peak in $T_{c}(P)$ near 40 GPa,
particularly in the susceptibility measurement, suggests a structural phase
transition at this pressure. Indeed, Feng \textit{et al.} \cite{feng1} predict
for hcp CaLi$_{2}$ significant lattice bifurcation at pressures $\gtrsim47$
GPa. A synchrotron radiation study at ambient temperature revealed no clear
evidence of any structural phase transitions to 20 GPa \cite{matsuoka3}; an
extension of these studies to higher pressures and lower temperatures would be
of particular interest since they would reveal whether the pressure-induced
superconductivity in CaLi$_{2}$ might be related to structural phase transitions.

In summary, samples of the hexagonal Laves phase compound CaLi$_{2}$ have been
prepared and characterized. At ambient pressure no superconductivity is found
above 1.10 K. Under pressures above 11 GPa, a superconducting transition is
observed in both the electrical resistivity and ac susceptibility at
temperatures as high as 13 K. The electrical resistivity increases by a
remarkable amount to 81 GPa, the highest pressure in this experiment. These
results thus lend support to the prediction by Feng \textit{et al.}
\cite{feng1} that, as for Ca and Li, the electronic properties in CaLi$_{2}$
become highly anomalous under pressure. High pressure studies would also be
very interesting on the nominally immiscible Li-Be system which has been
predicted to form a number of stable stoichiometric compounds under high
pressure which exhibit quasi-two-dimensional electronic structure
\cite{feng2}. Were one or more of these compounds to become superconducting,
higher values of $T_{c}$ than for CaLi$_{2}$ might be anticipated due to their
low molecular weight.\emph{\vspace{0.2cm}}

\noindent\textbf{Acknowledgments.} The authors acknowledge many stimulating
discussions with N. Ashcroft who also suggested these experiments. Thanks are
due J. Neumeier for helpful comments on the original manuscript. The research
visit of one of the authors (TM) at Washington University was made possible by
a Osaka University scholarship for short-term student dispatch program. The
authors gratefully acknowledge research support by the National Science
Foundation through Grant No. DMR-0703896.

\begin{center}
\bigskip{\LARGE Figure Captions}
\end{center}

\bigskip\ 

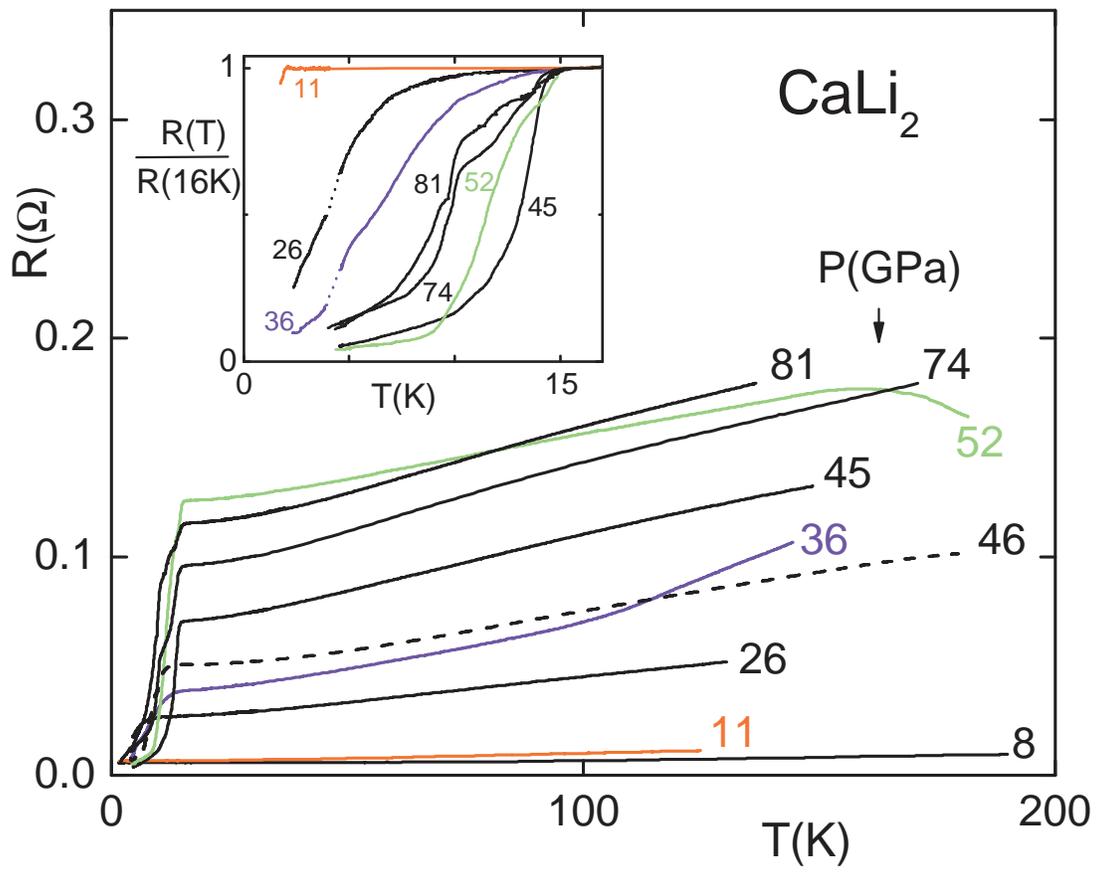
\noindent\textbf{Fig. 1. \ }(color online) Electrical resistance versus
temperature for CaLi$_{2}$ (sample A) at pressures 8, 11, 26, 36, 45, 52, 74,
81, 46, taken in that order. Inset shows data plotted as resistance normalized
at 16 K versus temperature to 17 K; dots bridge gap in data at 26 and 36
GPa.\bigskip

\noindent\textbf{Fig.\ 2.\ \ }(color online) Real part of ac susceptibility
versus temperature for CaLi$_{2}$ (sample B) for pressures 25, 35, 39, 43, 47,
54, taken in that order. In the inset, the superconducting transition for 39
GPa pressure shifts by 0.62 K to lower temperatures in 500 Oe dc magnetic
field.\bigskip

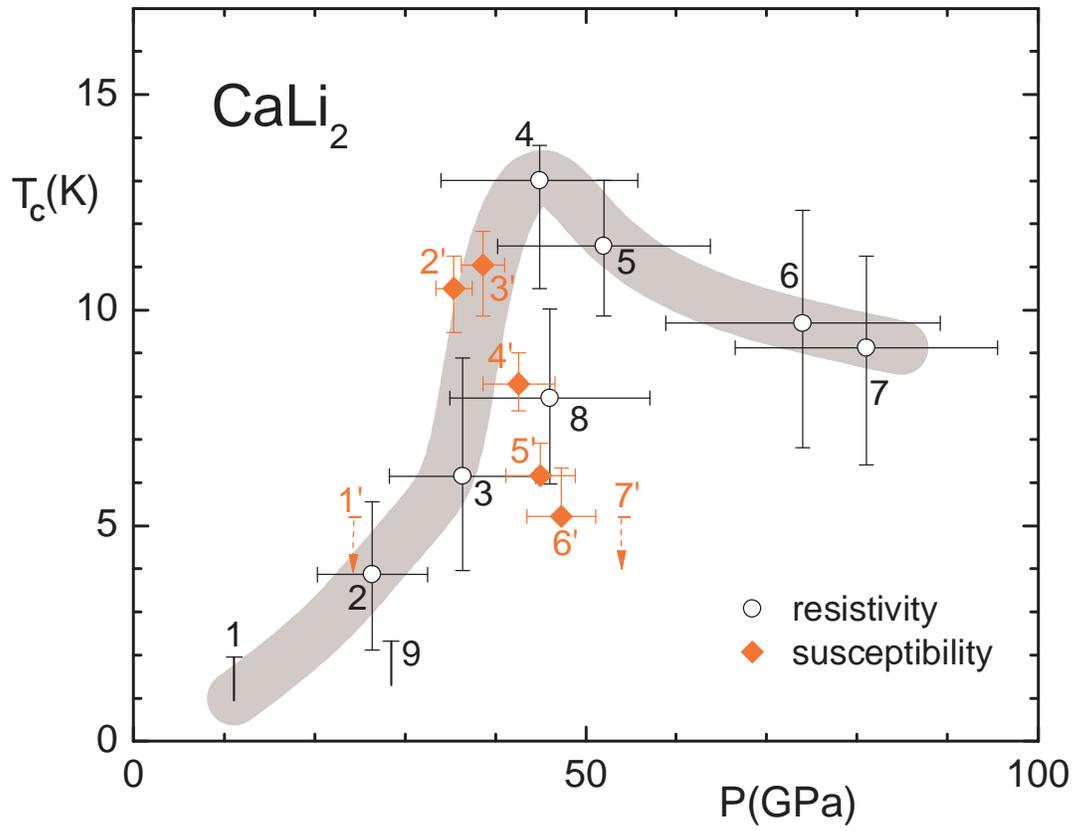
\noindent\textbf{Fig. 3. \ }(color online) Superconducting transition
temperature $T_{c}$ of CaLi$_{2}$ versus pressure from resistivity ($\circ$)
and ac susceptibility ($\blacklozenge$) measurements in Figs.~1 and 2. Value
of $T_{c}$ is determined from transition midpoint; vertical \textquotedblleft
error bars\textquotedblright\ give 20-80 transition width. Horizontal
\textquotedblleft error bars\textquotedblright\ reflect width of R$_{1}$ ruby
peak. Primed and unprimed numbers give order of measurement. Broad grey line
is guide to the eye.\newpage


\begin{thebibliography}{99}
\bibitem {experiment1}In a nonhydrostatic pressure cell an appreciable
pressure gradient exists across the sample, the pressure reaching its maximum
in the center. If the intrinsic $T_{c}(P)$-dependence passes through a
maximum, the region of the sample responsible for the onset temperature shifts
from the cell middle to the cell perimeter with increasing pressure which may
cause $T_{c}^{onset}(P)$ to \textquotedblleft hang up\textquotedblright\ at
its maximum value. In this situation the transition midpoint more closely
reproduces the intrinsic $T_{c}(P)$ - dependence.
\end{thebibliography}
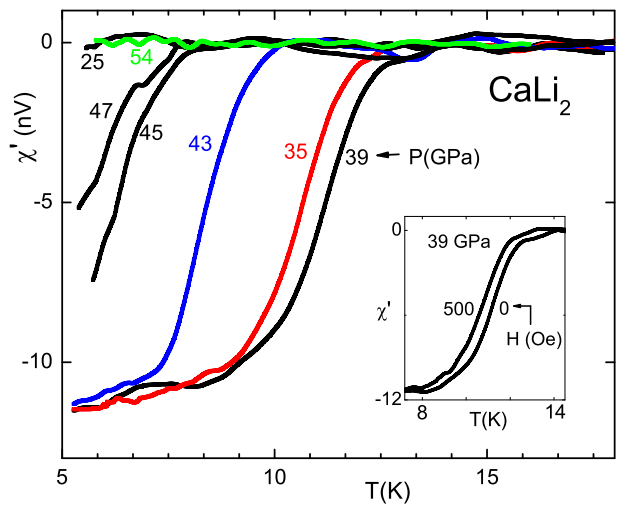
\end{document}